\documentclass[twocolumn,prl]{revtex4}
\usepackage{amsmath}
\usepackage{amssymb}

\newcommand{\ra}{\rangle}

\begin{document}

\title{The weak equivalence principle and the Dirac constant: A result from the holographic principle}

\author{Eiji Konishi}
\email{konishi.eiji.27c@kyoto-u.jp}
\address{Graduate School of Human and Environmental Studies, Kyoto University, Kyoto 606-8501, Japan}

\date{\today}

\begin{abstract}
In this article, based on a recent formularization of the holographic principle proposed and investigated by the present author, we show that the weak equivalence principle in general relativity is equivalent to the equivalence between two forms of the Dirac constant, that is, the action of the spin degree of freedom in the two-dimensional Hilbert space and the lower bound in the quantum mechanical uncertainty relations.
This result follows from an equation between the Euclidean and Lorentzian world-line actions of a massive particle divided by the Dirac constant, via the Wick rotation, by using the Euclidean and Lorentzian actions of a holographic tensor network, whose quantum state is classicalized by introducing the superselection rule.
\end{abstract}

\maketitle

The holographic principle \cite{Hol1,Hol2,Hol3} equates the externally observable degrees of freedom in the bulk spacetime with the amount of information relatively stored in the quantum pure state of quantum field theory (conformal field theory) defined on its codimension-one boundary spacetime without gravity.
Before the emergence of the holographic principle, the principles of unitary quantum mechanics (or quantum field theory) and general relativity had been considered to be completely distinct, with the former governing non-gravitational physics and the latter governing classical gravity.
The purpose of this article is to show that, in the holographic theory, there exists a fundamental equivalence between these two theories at the level of their principles.

For the concrete model of the holography, we consider the three-dimensional anti-de Sitter spacetime/two-dimensional conformal field theory (AdS$_3$/CFT$_2$) correspondence at the strong-coupling limit of the CFT$_2$, which is treated as a quantum many-body system of qubits \cite{AdSCFT1,AdSCFT2,Nastase,RT}.
We refer to the AdS$_3$ spacetime as the {\it bulk spacetime}, and the CFT$_2$ is defined on the boundary spacetime without gravity.
Based on advances pertaining to this correspondence in the holographic tensor network (HTN) theory \cite{Review1,Review2}, we replace the AdS$_3$ spacetime with a scale-invariant tensor network, that is, the multi-scale entanglement renormalization ansatz of the quantum mechanically entangled ground state of the boundary CFT$_2$ \cite{Swingle,MIH,Bao}.

The novelty of this article lies in the classicalization of this HTN by assuming the existence of a superselection rule (the one-qubit Pauli third matrix as the superselection rule operator\footnote{Effectively, this is the choice of a quantum reference frame.} \cite{BRS}) in the qubits ground state of the HTN \cite{EPL1,EPL2}.
Due to the superselection rule, the complete set of the qubits observables is restricted to an Abelian set of those that commute with the superselection rule operator, and then the quantum ground state becomes equivalent to a diagonal quantum mixed state with respect to this Abelian restricted set of the qubits observables \cite{WWW,Jauch,dEspagnat}: this means that the information stored in the quantum coherence (i.e., the off-diagonal part) of the ground state is completely lost.
This is the classicalized HTN (cHTN).
We denote the classicalized quantum ground state by $|\psi\ra=(|\psi\ra,{\cal A})$ for the Abelian restricted set ${\cal A}$ of the qubits observables by introducing the superselection rule.
We denote the Shannon entropy of this diagonal quantum mixed state $|\psi\ra$ in nats by $H[|\psi\ra]$.

Now, based on the holographic principle, we write down the actions of the cHTN of the ground state of the boundary CFT$_2$ in the Euclidean regime (spacetime) and the Lorentzian regime (spacetime) as
\begin{eqnarray}
I_E[|\psi\ra]&=&-\hbar_E H[|\psi\ra]\;,\label{eq:IE}\\
I_L[|\psi\ra]&=&-\hbar_L H[|\psi\ra]\;,\label{eq:IL}
\end{eqnarray}
respectively \cite{EPL2,JHAP3}.
The negative sign on the right-hand side of each equation indicates that information is lost in the boundary CFT$_2$ by the classicalization.
In other words, the bulk degrees of freedom are $-1$ (where the negative value indicates stochasticity) \cite{EPL2,JHAP1}.
Here, $-\hbar_E$ and $-\hbar_L$ are the actions of $-1$ degree of freedom (i.e., the actions of the {\it pixel}, that is, the classicalized disentangler \cite{EPL2}) in the cHTN in the Euclidean and the Lorentzian regimes, respectively, and we distinguish between $\hbar_E$ and $\hbar_L$ conceptually.
The physical roles of $\hbar_E$ and $\hbar_L$ are the action of the spin degree of freedom in the two-dimensional Hilbert space and the lower bound in the quantum mechanical uncertainty relations in the bulk spacetime, respectively \cite{JHAP7}.

First, we consider the Wick rotation
\begin{equation}
t_E=it_L\label{eq:t}
\end{equation}
between the Euclidean time $t_E$ (i.e., the imaginary time) and the Lorentzian time $t_L$ (i.e., the real time).
From the definitions (\ref{eq:IE}) and (\ref{eq:IL}) of the unit actions $-\hbar_E$ and $-\hbar_L$, respectively, this relation (\ref{eq:t}) is equivalent to the following relation:
\begin{equation}
\frac{M_E}{\hbar_E}=\frac{M_L}{\hbar_L}\;.\label{eq:frac}
\end{equation}
Here, the masses $M_E$ and $M_L$ are conceptually distinguished from each other and are respectively defined in the Euclidean world-line action $S_E$ and the Lorentzian world-line action $S_L$ of a massive particle in the bulk spacetime.
To show this equivalence between the relations (\ref{eq:t}) and (\ref{eq:frac}), we consider a massive particle in the cHTN.
Namely, this equivalence follows from the equivalence \cite{BPS}
\begin{equation}
\frac{S_E}{\hbar_E}=-i\frac{S_L}{\hbar_L}\biggr|_{t_L\to -it_E}\;.
\end{equation}
The reason why we consider the Euclidean world-line action $S_E$ divided by the Dirac constant $\hbar_E$ is that, for a massive particle in the cHTN, this quantity is the amount of information about {\it spin} events selected from the cHTN (locally, a statistical mixture of the two spin events \cite{JHAP1}) by the temporal increment of the Euclidean action of the particle \cite{JHAP3,JHAP1}.

Next, we consider the condition
\begin{equation}
\hbar_E=\hbar_L\;.\label{eq:QM}
\end{equation}
By assuming this value to be the Dirac constant $\hbar$, this condition indicates consistent derivation of (non-relativistic) path-integral unitary bulk quantum mechanics in the Lorentzian regime from imaginary-time path-integral in the Euclidean regime via the inverse Wick rotation \cite{JHAP1}.
From the relation (\ref{eq:frac}), this condition (\ref{eq:QM}) is equivalent to the following condition:
\begin{equation}
M_E=M_L\label{eq:GR}
\end{equation}
for an arbitrary massive particle in the bulk spacetime.
Here, $M_L$ appears in the rest energy $M_Lc^2$ as the energy uncertainty $\Delta E$ of the cHTN (\ref{eq:IL}) in the ground state.
So, $M_L$ is the inertial mass.
Specifically, in the Lorentzian regime, energy uncertainty $\Delta E$ of the cHTN is used in the physical interpretation of the on-shell equation of the Lorentzian action of the cHTN in the presence of a massive particle as the time--energy uncertainty relation of the cHTN in the ground state \cite{JHAP7,JHAP5}.
In the Euclidean regime, on the other hand, $M_E$ of a rest-massive particle linearly appears in an infinitesimal amount of information $d{\cal I} = d_{\tau_E} S_E/\hbar_E$ for the bulk imaginary proper time $\tau_E$, and $d{\cal I}$ is used as the direct source to derive the {\it gravitational} proper acceleration, which is a weak perturbation of {\it gravity}, in the background AdS$_3$ spacetime (the cHTN) at a distant site in the cHTN \cite{JHAP3}.
Specifically, the resultant physical Unruh proper acceleration ${\boldsymbol{a}}_E^U$, which is identified with the gravitational proper acceleration in the background AdS$_3$ spacetime, at a distant site in the cHTN has $M_Ed \tau_E$ as the source quantity in Gauss's theorem for it in the cHTN over an infinitesimal elapsed time $dt_E$ \cite{JHAP3}.
Thus, $M_E$ is the active gravitational mass.\footnote{Here, the active gravitational mass is the passive gravitational mass due to Eq. (21) in Ref. \cite{JHAP3}.}
Since ${\boldsymbol{a}}_E^U$ is the Unruh proper acceleration, it is introduced without the inertial mass.
These identities (\ref{eq:GR}) for the massive particles in the bulk spacetime therefore correspond to the weak equivalence principle in general relativity, namely, the equivalence between the inertial mass $M_L$ and the gravitational mass $M_E$ in the Lorentzian and the Euclidean bulk spacetimes, respectively.

This equivalence between the relations (\ref{eq:QM}) and (\ref{eq:GR}) is the equivalence between the principle for the consistency of unitary bulk quantum mechanics in the framework of the cHTN \cite{EPL2,JHAP1} and the weak equivalence principle in general relativity.
In conclusion, this result suggests that quantum mechanics and general relativity are two sides of the same coin at the level of their principles.

\end{document}